\documentclass[aps,prl,twocolumn,showpacs,lettersize,superscriptaddress]{revtex4-1}

\usepackage{hyperref}
\usepackage{amsmath}
\usepackage{amssymb}
\usepackage{color}
\usepackage{subfigure}
\usepackage{graphicx}
\usepackage[varg]{txfonts}
\usepackage{natbib}

\begin{document}

\title{Observation of vortex dipoles in an oblate {B}ose-{E}instein condensate}

\author{T. W. Neely}
\author{E. C. Samson}
\affiliation{College of Optical Sciences, University of Arizona,
Tucson, AZ 85721, USA}
\author{A. S. Bradley}
\affiliation{Jack Dodd Centre for Quantum Technology, Department of Physics, University of Otago, P. O. Box 56, Dunedin, New Zealand}
\author{M. J. Davis}
\affiliation{The University of Queensland, School of Mathematics and Physics, ARC Centre of Excellence for Quantum-Atom Optics, Qld 4072, Australia}
\author{B. P. Anderson}  
\affiliation{College of Optical Sciences, University of Arizona,
Tucson, AZ 85721, USA}
\affiliation{Department of Physics, University of Arizona,
Tucson, AZ 85721, USA}

\date{December 17, 2009}

\begin{abstract}
We report experimental observations and numerical simulations of the formation, dynamics, and lifetimes of single and multiply charged quantized vortex dipoles in highly oblate dilute-gas Bose-Einstein condensates (BECs).  We nucleate pairs of vortices of opposite charge (vortex dipoles)  by forcing superfluid flow around a repulsive gaussian obstacle within the BEC.  By controlling the flow velocity we determine the critical velocity for the nucleation of a single vortex dipole, with excellent agreement between experimental and numerical results.  We present measurements of vortex dipole dynamics, finding that the vortex cores of opposite charge can exist for many seconds and that annihilation is inhibited in our highly oblate trap geometry.  For sufficiently rapid flow velocities  we find that clusters of like-charge vortices aggregate into long-lived dipolar flow structures.
\end{abstract}


\maketitle

Vortex dipoles consist of a bound pair of vortices of opposite circulation and may exist in both classical and quantum fluids. Although single vortices carry angular momentum, vortex dipoles can be considered as basic topological structures that carry \emph{linear} momentum~\cite{Voropayev1994a} in stratified or two-dimensional fluids. Vortex dipoles are widespread in classical fluid flows, appearing for example in ocean currents~\cite{Ginzburg1984a} and soap films~\cite{Couder1986a}, and have been described as the primary vortex structures in two-dimensional chaotic flows~\cite{Voropayev1994a}.  In superfluids, the roles of quantized vortex dipoles appear less well established.  Given the prevalence of vortices and antivortices in superfluid turbulence~\cite{QVDSF,Kobayashi2005a,Horng2008a}, the Berezinskii-Kosterlitz-Thouless (BKT) transition~\cite{Hadzibabic2006a}, and phase transition dynamics~\cite{Zurek1985,anglin,svistunov,Weiler2008a}, a quantitative study of vortex dipoles will contribute to a broader and deeper understanding of superfluid phenomena.  The realization of vortex dipoles in dilute Bose-Einstein condensates (BECs) is especially significant as BECs provide a clean testing ground for the microscopic physics of superfluid vortices~\cite{Fetter2001a,Kevrekidis2004a,Keverkidis2008a}.  In this paper we present an experimental and numerical study of the formation, dynamics, and lifetimes of single and multiply charged vortex dipoles in highly oblate BECs.

Numerical simulations based on the Gross-Pitaevskii equation (GPE) have shown that vortex dipoles are nucleated when a superfluid moves past an impurity faster than a critical velocity, above which vortex shedding induces a drag force~\cite{Frisch1992a,Winiecki1999a}. Vortex shedding is therefore believed to be a mechanism for the breakdown of superfluidity~\cite{Jackson1998a,Winiecki2000a}. Experimental studies of periodic stirring of a BEC with a laser beam have measured a critical velocity for the onset of heating and a drag force on superfluid flow~\cite{Raman1999a,Onofrio2000a}, and vortex phase singularities have been observed in the wake of a moving laser beam~\cite{Inouye2001a,Jackson2000b}. However, a microscopic picture of vortex dipole formation and the ensuing dynamics has not been established experimentally. In the work reported here, single vortex dipoles are deterministically nucleated by causing a highly oblate, harmonically trapped BEC to move past a repulsive obstacle.  We measure a critical velocity for vortex dipole shedding, and find good agreement with numerical simulations and earlier theory~\cite{Crescimanno2000a}. Experimentally, the nucleation process exhibits a high degree of coherence and stability, allowing us to map out the orbital dynamics of a vortex dipole. We find that vortex dipoles can survive for many seconds in the BEC without self-annihilation.  We also provide evidence for the formation of multiply charged vortex dipoles.

\begin{figure}
\includegraphics[width=87mm]{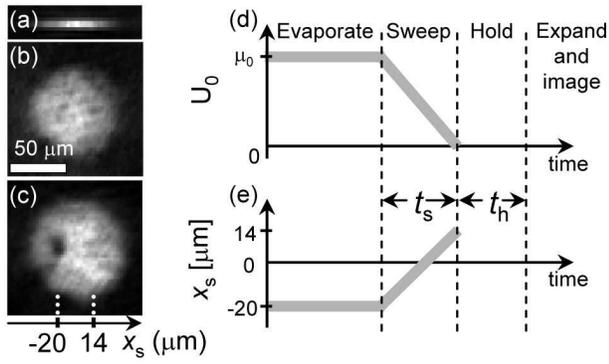}
\caption{Illustration of the BEC initial state and the experimental sequence.  (a) Side-view phase-contrast image and
(b) axial absorption image of a BEC in the highly oblate harmonic trap in the absence of the obstacle. Lighter shades indicate higher column densities integrated along the line of sight.  Our axial and radial trapping frequencies are $\omega_z = 2\pi \times (90\; \mathrm{Hz})$ and $\omega_r = 2\pi \times (8\; \mathrm{Hz})$, respectively.    (c) BEC initial state with the obstacle located at $x_s = -20\, \mu$m relative to the BEC center.  (d,e) The maximum repulsive potential energy of the obstacle is $U_0 \approx 1.2 \mu_0$ (where $\mu_0 \sim 8\hbar\omega_z$ is the BEC chemical potential) and is held constant during evaporative cooling.  It is ramped down linearly as the trap translates; relative to the trap center, the beam moves from position $x_s = -20$ $\mu$m to $x_s = 14$ $\mu$m over a variable sweep time $t_s$.  The BEC is then held in the harmonic trap for a variable time $t_h$ prior to expansion and absorption imaging.}
\label{fig1}
\end{figure}

\begin{figure}[th]
\includegraphics[width=87mm]{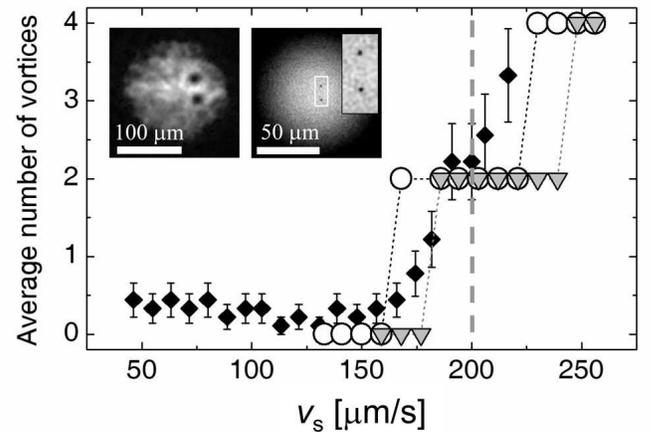}
\caption{Average number of vortex cores observed for a range of translation velocities $v_s$ with $N_c \sim 2\times 10^6$ atoms in the BEC.  Experimental data points (black diamonds) are averages of 10 runs each, with error bars showing the standard deviation of the observations. Numerical data for $N_c = 2\times 10^6$ at a system temperature of $T=52$ nK, corresponding to the experimental conditions, are indicated by triangles joined with dotted lines.  Fewer atoms and lower temperatures reduce the critical velocity; as an example, open circles show the results of numerical simulations using  $N_c = 1.3\times 10^6$ and $T=0$.  For  $v_s \lesssim 170\,\,\mu {\rm m}/$s, remnant vortices arising spontaneously during condensate formation (see Ref.~\cite{Weiler2008a}) account for the baseline of $\sim 0.3$.   For $v_s \sim 190\,\,\mu{\rm m} /$s, the inset images show a single pair of vortices having formed in the experiment (left) and simulations (right).  Because the vortex core diameters are well below our imaging resolution, the BEC is expanded prior to imaging.  Similarly, the vortex cores in the unexpanded numerical image are barely visible; a 10-$\mu$m-wide inset provides a magnified scale for the core size.  Above $v_s \sim200$ $\mu{\rm m}/$s, multiply-charged vortex dipoles are observed.  The critical velocity calculated using the methods of Ref.~\cite{Crescimanno2000a} is indicated by the vertical dashed line.}
\label{fig2}
\end{figure}

The creation of BECs in our lab is described elsewhere \cite{Scherer2007a,Weiler2008a,epaps}.  In the experiments reported here, we begin with a BEC of $2\times10^6$ atoms in a highly oblate harmonic trap.  Our axially symmetric trap is created by combining a red-detuned laser light-sheet trapping potential with a magnetic trapping potential, producing a BEC with an 11:1 aspect ratio and a Thomas-Fermi radius of 52 $\mu$m radially, as shown in Fig.~\ref{fig1}(a,b).  The BECs are additionally penetrated by a focused blue-detuned laser beam that serves as a repulsive obstacle; the beam has a Gaussian 1/$e^2$ radius of 10 $\mu$m and is initially located $20$ $\mu$m to the left of the minimum of the harmonic trap as shown in  Fig.~\ref{fig1}(c).  To nucleate vortices we translate the harmonic potential in the horizontal ($x$) direction at a constant velocity until the obstacle ends up $14$ $\mu$m to the right of the harmonic trap minimum.   At the same time, the height of the obstacle is linearly ramped to zero as shown in Fig.~\ref{fig1}(d,e), allowing us to gently create a vortex dipole that is unaffected by the presence of an obstacle or by heating due to moving the obstacle through the edges of the BEC where the local speed of sound is small. For our conditions, the healing length at trap center is $\sim0.3$ $\mu$m, which is approximately the size of vortex cores in our trapped BECs. After a subsequent variable hold time $t_h$ we remove the trapping potential and expand the BEC for imaging, causing the vortex cores to expand relative to the condensate radius such that they are optically resolvable. An example axial absorption image is shown in the left-most inset image of Fig.~\ref{fig2}.

%%%%%%%%%%%
%FIGURE 1 %
%%%%%%%%%%%

%%%%%%%%%%%
%FIGURE 2 %
%%%%%%%%%%%

In Fig.~\ref{fig2} we plot the average number of vortices observed as a function of the trap translation velocity $v_s$. In our experimental procedure we find a $\sim 30$\% likelihood of a single vortex occurring during condensate formation even prior to trap translation; see Ref.~\cite{Weiler2008a} for further details.  This gives a non-zero probability of observing a single vortex for the lowest translation velocities in Fig.~\ref{fig2} even when flow without drag is expected.  The results of zero-temperature and finite-temperature c-field numerical simulations~\cite{Blakie2008a} are also shown --- for simulation details see Ref.~\cite{epaps}. There is good agreement between simulation and experimental results, and we identify a critical velocity $v_c$ for vortex dipole formation between 170 $\mu$m$/$s and 190 $\mu$m$/$s for $N_c = 2\times10^6$ atoms and temperature $T=52$ nK.
Recently, Crescimanno {\em et al.}~\cite{Crescimanno2000a} have calculated the critical velocity for vortex dipole formation in a 2D BEC in the Thomas-Fermi regime. By using the nonlinearity and chemical potential of our 3D system in their 2D expression, we estimate a critical velocity of 200 $\mu$m$/$s.  For our conditions, the maximum speed of sound at the trap center is calculated to be $c \sim$ 1700  $\mu$m$/$s; our measurements show that $v_c \sim 0.1 c$, consistent with previous measurements of a critical velocity for the onset of a drag force~\cite{Onofrio2000a}.

%%%%%%%%%%%
%FIGURE 3 %
%%%%%%%%%%%
\begin{figure*}[!htb]
\includegraphics[width=169mm]{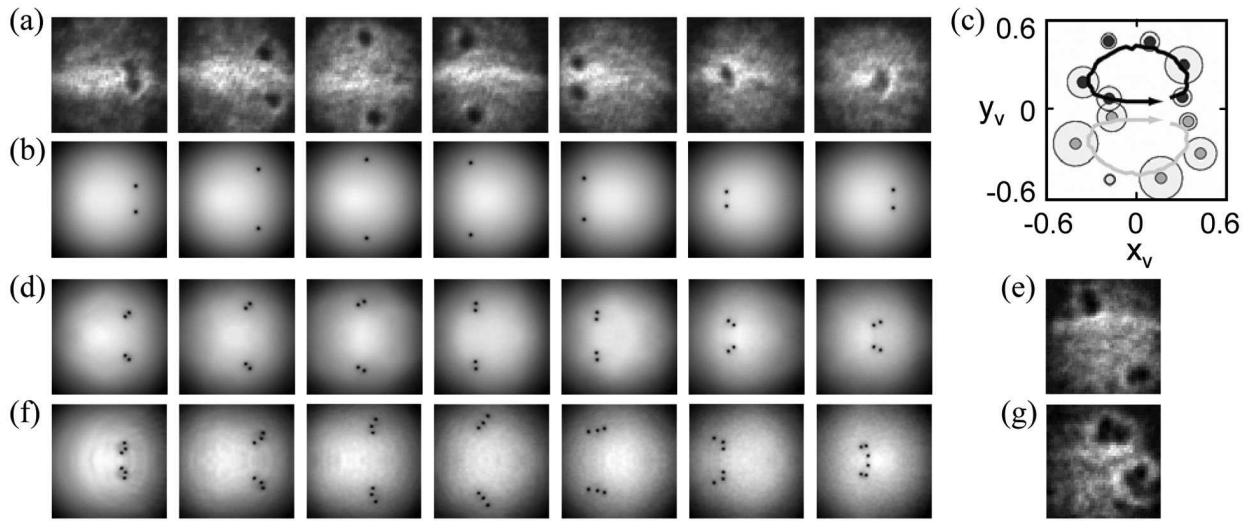}
\caption{Sequences of images showing the first orbit of vortex dipole dynamics.  (a) Back-to-back expansion images~\cite{sequencenote1} from the experiment with 200 ms of successive hold time between the 180-$\mu$m-square images.  This data sequence was taken using an obstacle height 3.15 times larger than that used for the data of Fig.~\ref{fig2}, as we  find this gives the highest degree of repeatability and the least sensitivity to beam displacement.  (b) 62-$\mu$m-square images from numerical data obtained for conditions similar to the data of sequence (a), but for a temperature of $T=0$.  The apparent vortex core size is smaller in the simulations as we show in-trap data rather than expanded BECs. (c) Black and dark gray small circles show average positions $x_v$ and $y_v$ of each of the two vortices from 5 sequences of experimental data, each sequence using a procedure identical to that of sequence (a).  The larger circle around each average position point represents the standard deviation of the vortex positions at that specific hold time, and is calculated from the 5 images obtained at that time step.  A continuous dipole trajectory from sequence (b) is re-scaled to the Thomas-Fermi radius of the expanded experimental images, and superimposed as solid lines on the experimental data;  no further adjustments are made for this comparison.  (d) Similar to the sequence of (b), but for a faster translation velocity with which a doubly charged vortex dipole is formed.  Images are spaced in time by 120 ms as the vortices orbit more quickly.  (e) An experimental image in which a doubly charged vortex is formed; see text for further explanation. (f,g) Similar to (d,e) but for a triply charged dipole and 100 ms between images in (f).}
\label{fig3}
\end{figure*}

In an axi-symmetric trap such as ours, a vortex dipole coincides with a meta-stable state of superfluid flow with potentially long lifetimes~\cite{Crasovan2003a,Mottonen2005a,Li2008a}. The vortices exhibit periodic orbital motion, a 2D analogue of the dynamics of a single vortex ring~\cite{Jackson2000b,Anderson2001a}. To observe these dynamics we nucleate a single vortex dipole and hold the BEC for variable time $t_h$ prior to expansion and imaging, with results shown in  Fig.~\ref{fig3}(a).
The repeatability and coherence of the vortex nucleation process is clear: in back-to-back images with increasing $t_h$, the vortex positions and orbital dynamics can be followed and the dipolar nature of the superfluid flow is microscopically determined.  These measurements also determine the direction of superfluid circulation about the vortex cores, analogous to the case of single vortices~\cite{Anderson2000a}: the image sequence shows counter-clockwise fluid circulation about the vortex core in the upper half of the BEC and clockwise circulation in the lower half.  The orbital dynamics were also examined in zero-temperature GPE simulations, as shown in Fig.~\ref{fig3}(b)~\cite{epaps}, and the experimental and numerical data are in good agreement regarding vortex dipole trajectories, as shown in Fig.~\ref{fig3}(c).  The lifetime of a single vortex dipole is much longer than the first orbital period of $\sim$1.2 s (see below), although after the first orbit the vortex trajectories become less repeatable from shot-to-shot. However, it is the large-scale flow pattern of the first orbit that is perhaps most indicative of the qualitative connection between 2D superfluid and classical dipolar fluid flows.

For trap translation velocities well above $v_c$ we observe the nucleation of \emph{multiply-charged} vortex dipoles both experimentally and numerically, as shown in Fig.~\ref{fig3}(d-g). Viewed on a coarse scale the ensuing dynamics are consistent with that of a dipole comprised of a highly charged vortex and antivortex. On a fine scale, particularly in numerical data, we see loose aggregations of singly quantized vortices with the same circulation at the two loci of vorticity in the dipolar flow. In the experimental images obtained at higher sweep speeds, many individual vortices are often not clearly resolvable for the short hold times shown.  Nevertheless, the data resemble characteristics of highly charged dipoles and suggest the formation of many vortices because (i) the apparent vortex core sizes become larger, (ii) the orbital time period for the dipole structure is shorter, as expected for higher numbers of cores and faster superfluid flow, and (iii) multiple individual vortex cores are observable for longer hold times.  Although we have not performed an exhaustive analysis of these states, we present these results to bring attention to these interesting metastable superfluid vortex structures.

\begin{figure}[t]
\includegraphics[width=87mm]{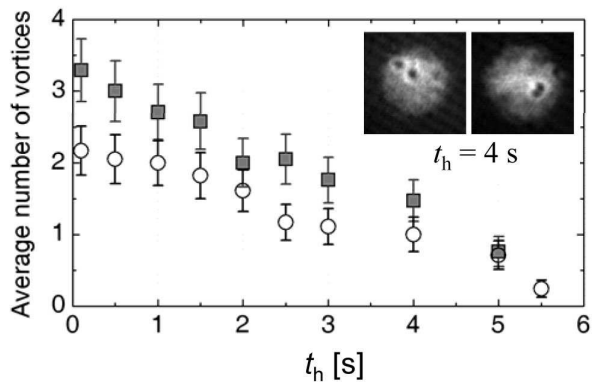}
\caption{Number of vortices remaining after dipole nucleation as a function of hold time $t_h$, averaged over 17 realizations per data point.  The circles show conditions for which a singly charged dipole is created, while the squares show data from a faster sweep where four cores (a doubly charged dipole) preferably occur.  The error bars represent statistical uncertainty (as in Fig.~\ref{fig2}) rather than counting uncertainty; however, for the doubly charged dipole case there is additional uncertainty in counting vortex cores because we do not always resolve 4 well-defined cores at the earlier hold times.  The vortex cores can clearly persist for times much longer than the dipole orbital period of $\sim$1.2 s for a singly quantized dipole, and $\sim0.8$ s for a doubly quantized dipole.  The inset figures show experimental images of singly quantized vortices seen after 4 seconds of hold time in the harmonic trap.}
\label{fig4}
\end{figure}

While it is often assumed that in a finite-temperature environment, vortices of opposite circulation will attract and annihilate each other at close distances, this is not necessarily the case: vortices may approach each other so closely that they appear to coalesce --- see for example the sixth image of Fig.~\ref{fig3}(a) with $t_h=1$ s --- and yet still move away from each other after the encounter. In Fig.~\ref{fig4} we show measurements of the average number of vortices observed with various hold times after nucleating a vortex dipole, from which we conclude that singly and doubly quantized vortex dipoles may exhibit lifetimes of many seconds, much longer than a single orbital period. With such a strong trap asymmetry, the vortex lines are relatively impervious to bending~\cite{Bretin2003a} and tilting~\cite{Haljan2001a}, and annihilation is suppressed because vortex crossings and reconnections are inhibited.

%%%%%%%%%%%
%FIGURE 4 %
%%%%%%%%%%%

Although vortex shedding is the microscopic mechanism for the breakdown of superfluid flow and the onset of a drag force, the nucleation of a vortex dipole does not imply immediate superfluid \emph{heating} since dipoles are coherent structures that are metastable and do not immediately decay into phonons. We estimate the maximum energy of a vortex dipole in the BEC is $k_B \times 0.45$ nK/atom for our system~\cite{Raman1999a}, so that at $T=52$ nK the temperature would increase by less than 0.5\% upon self-annihilation of a single dipole. This temperature change would be very difficult to measure, and microscopic observations of vortex dipole formation and dynamics are therefore complementary to heating and drag force observations.

In conclusion, we have demonstrated controlled, coherent nucleation of single vortex dipoles in oblate BECs. The dipole dynamics reveal the topological charges of the vortices and show that the dipole is a long-lived excitation of superfluid flow. Sufficiently rapid translation of the harmonic trap causes vortices with identical charge to aggregate into highly charged dipolar vortex structures that exhibit orbital dynamics and long lifetimes analogous to singly charged vortex dipoles. This suggests that both dipole structures and macro-vortex states are readily accessible in highly oblate and effectively two-dimensional superfluids.

%%%%%%%%%%%%%%%%%%%%%%%%%%%%%%%%%%%%%%%%%%%%%%%%%%%%%%%%%%%%%%%%%%%%%%%%%%%%%%%%%%%%%%%%%%%%%%%%%
% Acknowledgements
%%%%%%%%%%%%%%%%%%%%%%%%%%%%%%%%%%%%%%%%%%%%%%%%%%%%%%%%%%%%%%%%%%%%%%%%%%%%%%%%%%%%%%%%%%%%%%%%%

\begin{acknowledgments}
We thank David Roberts for many helpful discussions.  The authors acknowledge funding from  the US National Science Foundation grant number PHY-0855467, the US Army Research Office, the New Zealand Foundation for Research, Science, and Technology contract UOOX0801, and the Australian Research Council Centre of Excellence program.
\end{acknowledgments}

%%%%%%%%%%%%%%%%%%%%%%%%%%%%%%%%%%%%%%%%%%%%%%%%%%%%%%%%%%%%%%%%%%%%%%%%%%%%%%%%%%%%%%%%%%%%%%%%%
%%%%%%%%%%%%%%%%%%%%%%%%%%%%%%%%%%%%%%%%%%%%%%%%%%%%%%%%%%%%%%%%%%%%%%%%%%%%%%%%%%%%%%%%%%%%%%%%%
% END OF PAPER
%%%%%%%%%%%%%%%%%%%%%%%%%%%%%%%%%%%%%%%%%%%%%%%%%%%%%%%%%%%%%%%%%%%%%%%%%%%%%%%%%%%%%%%%%%%%%%%%%
%%%%%%%%%%%%%%%%%%%%%%%%%%%%%%%%%%%%%%%%%%%%%%%%%%%%%%%%%%%%%%%%%%%%%%%%%%%%%%%%%%%%%%%%%%%%%%%%%

%%%%%%%%%%%%%%%%%%%%%%%%%%%%%%%%%%%%%%%%%%%%%%%%%%%%%%%%%%%%%%%%%%%%%%%%%%%%%%%%%%%%%%%%%%%%%%%%%
%%%%%%%%%%%%%%%%%%%%%%%%%%%%%%%%%%%%%%%%%%%%%%%%%%%%%%%%%%%%%%%%%%%%%%%%%%%%%%%%%%%%%%%%%%%%%%%%%
% References
%%%%%%%%%%%%%%%%%%%%%%%%%%%%%%%%%%%%%%%%%%%%%%%%%%%%%%%%%%%%%%%%%%%%%%%%%%%%%%%%%%%%%%%%%%%%%%%%%
%%%%%%%%%%%%%%%%%%%%%%%%%%%%%%%%%%%%%%%%%%%%%%%%%%%%%%%%%%%%%%%%%%%%%%%%%%%%%%%%%%%%%%%%%%%%%%%%%

%\begin{thebibliography}
%\bibliographystyle{apsrev4-1}
%to generate a bibtex file:
%\bibliography{dipoleRefs5.bib}

\begin{thebibliography}{10}
\bibitem{sequencenote1}%
  \BibitemOpen
  \bibinfo {note} {The final image shown was taken as part of a different
  sequence, but is included here to more fully illustrate the dipole dynamics.}%
  \bibAnnoteFile{Stop}{sequencenote1}%
\end{thebibliography}
%\end{thebibliography}

%Merlin.mbs v4.21 2009-07-09.

%

%%%%%%%%%%%%%%%%%%%%%%%%%%%%%%%%%%%%%%%%%%%%%%%%%%%%%%%%%%%%%%%%%%%%%%%%%%%%%%%%%%%%%%%%%%%%%%%%%%%%
%
%
\end{document}

\title{\large AUXILIARY MATERIAL FOR: Observation of vortex dipoles in an oblate {B}ose-{E}instein condensate}

\author{T. W. Neely}
\author{E. C. Samson}
\affiliation{College of Optical Sciences, University of Arizona,
Tucson, AZ 85721, USA}
\author{A. S. Bradley}
\affiliation{Jack Dodd Centre for Quantum Technology, Department of Physics, University of Otago, P. O. Box 56, Dunedin, New Zealand}
\author{M. J. Davis}
\affiliation{The University of Queensland, School of Mathematics and Physics, ARC Centre of Excellence for Quantum-Atom Optics, Qld 4072, Australia}
\author{B. P. Anderson}  
\affiliation{College of Optical Sciences, University of Arizona,
Tucson, AZ 85721, USA}
\affiliation{Department of Physics, University of Arizona,
Tucson, AZ 85721, USA}

\date{\today}

\begin{abstract}
Here we provide auxiliary details regarding our experimental methods and theoretical techniques for studying and modeling vortex dipoles in Bose-Einstein condensates, as reported in the main paper~[T. W. Neely \emph{et al.}].
\end{abstract}

\maketitle

{\em Experimental methods:} In previous work~\cite{Scherer2007a,Weiler2008a}, we described our methods to create $^{87}$Rb BECs in the $F=1,\,m_F=\,-1$ state in a time-averaged orbiting potential (TOP) trap, which was combined with optical dipole potentials from a blue-detuned laser. To study BECs in a highly oblate geometry, we further compress the trap in the vertical ($z$) dimension by adding a horizontally propagating, 1-W, sheet of red-detuned laser light of wavelength 1090 nm.  The resulting trap has radial and axial trapping frequencies of $(\omega_r,\omega_z)=2\pi\times(8,90)$ Hz, with the radial frequency dominantly determined by the TOP trap portion of the potential.  Radio-frequency evaporative cooling in this highly oblate harmonic trap achieves condensation at a critical temperature of $T_c \sim 90$ nK, and produces a BEC of $N_c = 2\pm0.5\times 10^6 $ atoms at a final temperature of $T\sim$ 52 nK.  The BEC chemical potential in the harmonic trap is $\mu_0~\sim~8\hbar\omega_z$, placing our BECs well within the three-dimensional (3D) regime. Radial and axial BEC images are shown in Fig.~1(a,b) of the main paper.

The repulsive gaussian obstacle used to nucleate vortex dipoles is formed by a blue-detuned 660-nm laser beam propagating along the axial ($z$) direction of the trap. This beam has a Gaussian 1/$e^2$ radius of 10 $\mu$m at the BEC, much smaller than the 52 $\mu$m Thomas-Fermi radius of the BEC, and the height of the repulsive potential has an energy of approximately $1.2\mu_0$.  The beam is turned on prior to condensation, and is aligned 20 $\pm$ 3 $\mu$m to the left of the trap center as shown in Fig.~1(c) of the main paper.  After the BEC forms in this annular trap, the harmonic potential is translated in the horizontal ($x$) direction at a constant velocity until the beam position ends up 14 $\mu$m to the right of the trap center and the translation then stops.  Trap translation is accomplished by changing a magnetic bias field, a process that we find to be more repeatable and stable than moving the beam with piezo-electric transducers. To initiate trap translation, the position of the trap center is smoothly accelerated over $\sim$5 ms before reaching a constant translation velocity for the duration of the translation stage.  This brief acceleration stage ensures that residual ``sloshing'' of the BEC due to excitation of center-of-mass oscillations is minimized; experimentally we find the amplitude of such oscillatory motion to be less than 1 $\mu$m.  Coincident with trap displacement, the beam intensity is ramped to zero as shown in Fig.~1(d,e) of the main paper.  After a variable hold time $t_h$, we enhance vortex visibility by expanding the BEC radially for 15 ms in the presence of the laser sheet with the TOP trap off, followed by free expansion for an additional 41 ms. An example axial absorption image is shown in the left inset image of Fig.~2 of the main paper.

\emph{Simulations for $T=0$:}
At zero temperature we simulate the experiment using a 3D Gross-Pitaevskii equation (GPE).  We find an initial state for the combined harmonic-plus-obstacle trap with $2\times 10^6$ atoms by integrating the GPE in imaginary time.  The potential due to the obstacle laser beam is $U_{{\rm S}}(x,y,t)= U_0(t)\exp{\{-2([x-x_0(t)]^2 + y^2)/w_0^2\}}$, where $U_0(0)=93$ $\hbar\omega_r=k_B\times 35$ nK $ \sim \mu_0 $, and $w_0=10$ $\mu {\rm m}$. We assume that $U_{{\rm S}}$ is constant over the axial position coordinate $z$, and $x$ and $y$ are radial position coordinates with $(x,y,z)=(0,0,0)$ corresponding to the harmonic trap center.  We model the vortex dipole nucleation procedure by translating the obstacle while leaving the harmonic trap center fixed at $(x,y,z)=(0,0,0)$. The obstacle laser beam moves along the $x$-axis from position $x_0(0)=-20\,\mu{\rm m}$ to $x_0(t_s)=14\,\mu{\rm m}$ during time $t_s$ and at a constant velocity $v_s = (34\, \mu{\rm m})/t_s$.  As in the experiment, the laser intensity, and thus $U_0(t)$, linearly ramps to zero over time $t_s$.  We then continue the simulations to observe the $T=0$ dynamics of the condensate and any vortices that were nucleated.  We vary $v_s$ over the range 133 to $256\,\mu{\rm m}/$s in steps of $8\,\mu{\rm m}/$s.

We emphasize that the simulations described above do not directly correspond to the experimental sequence, where the harmonic trap is translated rather than the obstacle. While the appropriate timescale for strictly adiabatic dynamics is $\omega_r^{-1}=20{\rm ms}\ll t_a$, a shorter acceleration stage is experimentally found to be suitable for the trap motion used, as mentioned above.   Thus our simulations are equivalent to applying an instantaneous Gallilean boost to the condensate at $t=0$ so that the condensate is stationary in the translating frame.  We refer to this as the ``adiabatic'' regime of trap acceleration.

We have performed additional simulations to investigate the ``impulse'' regime where the harmonic trap is instantly accelerated to the sweep velocity.  In this case we perform the simulation in the frame co-moving with the trap --- the transformation to this frame results in an initial phase gradient across the BEC causing it to initially move with the obstacle.  This procedure excites a center-of-mass oscillation, typically of amplitude 10 $\mu$m or more near the critical velocity, much larger than observed in the experiment.  These numerical observations thus suggest that the experiment is far closer to the ``adiabatic'' regime results presented in the main paper.  We note that simulations in the impulse regime also show a significantly lower critical velocity ($\sim 80$ $\mu$m/s), and multiple vortex pairs are much more readily nucleated at lower trap translation velocities.  This appears to be due to relatively high velocities between the obstacle and the oscillatory motion of the BEC center-of-mass (due to instantaneous trap acceleration); factors other than the constant velocity of the trap must therefore be considered.
A complete model of the experiment would need to incorporate details of the trap acceleration that were not obtained.  However, given the excellent agreement between the adiabatic regime of simulations and the full set of experimental results, including the experimental observations showing minimal BEC sloshing after trap acceleration, we conclude that our simulation methods are appropriate for comparisons with the experimental data.

\emph{Simulations for $T>0$:} While the experiments are performed at the lowest attainable temperatures, there is still a non-negligible thermal fraction present.  For this reason we have performed simulations of the vortex nucleation process using c-field simulation techniques~\cite{Blakie2008a}.  Broadly this is an extension of the Gross-Pitaevskii equation to finite temperatures where the lowest energy modes of the system have a large thermal occupation and it is reasonable to treat them classically.

Our procedure is as follows.  We find an initial finite-temperature state for the system by evolving the stochastic Gross-Pitaevskii equation \cite{Gardiner2003b,Blakie2008a} to thermal equilibrium for an input chemical potential of k$_B \times 37$ nK corresponding to a condensate number of $2.0 \times 10^6$ atoms, and a temperature of 52 nK as measured in the experiment.  Once we have generated a number of initial states for a given temperature, we apply the same simulation procedure as described above for the zero-temperature GPE.   We perform our simulations in the harmonic oscillator basis, and implement a basis cutoff at 130 $\hbar \omega_r$.  This is necessary in order to accurately account for the modification of the condensate wave function by the obstacle beam --- in doing so we include a rather large range of excited states that are not highly occupied in the classical field $\psi$, hence violating the usual validity condition for classical field simulations.  However, this is unavoidable, and we expect that it will not have significant effects on the nucleation of the vortex dipoles, although it may affect their subsequent dynamics.

%\bibliography{dipoleRefs5}